\documentclass[aip, amsmath,amssymb,reprint,]{revtex4-2}

\usepackage{comment}

\usepackage{graphicx}
\usepackage{dcolumn}
\usepackage{bm}

\usepackage[utf8]{inputenc}
\usepackage[T1]{fontenc}
\usepackage{mathptmx}
\usepackage{etoolbox}

\usepackage[dvipsnames]{xcolor}
\usepackage{xspace}

\usepackage{amsfonts}
\usepackage{comment}
\usepackage{hyperref}
\usepackage[font=small]{caption}
\usepackage{subcaption}
\usepackage{units}
\usepackage{cancel}

\begin{document}

\title{
Statistical Mechanics of Thermostatically Controlled Multi-Zone Buildings
}

\author{Lucas Fuentes Valenzuela}
\affiliation{Department of Electrical Engineering, Stanford University}
\author{Lindell Williams}
\affiliation{Department of Physics, Cornell University} 
\affiliation{Program in Applied Mathematics  \& Department of Mathematics, University of Arizona}
\author{Michael Chertkov}
\affiliation{Program in Applied Mathematics  \& Department of Mathematics, University of Arizona}

\email{lucasfv@stanford.edu; chertkov@arizona.edu}

\date{\today}

\begin{abstract}

We study the collective phenomena and constraints associated with the aggregation of individual cooling units from a statistical mechanics perspective. These units are modelled as Thermostatically Controlled Loads (TCLs) and represent zones in a large commercial or residential building. Their energy input is centralized and controlled by a collective unit -- the Air Handling Unit (AHU) --- delivering cool air to all TCLs, thereby coupling them together. Aiming to identify representative qualitative features of the AHU-to-TCL coupling, we build a realistic but also sufficiently simple model and analyze it in two distinct regimes: the Constant Supply Temperature (CST) and the Constant Power Input (CPI) regimes.  In both cases, we center our analysis on the relaxation dynamics of individual TCL temperatures to a statistically steady state. We observe that while the dynamics are relatively fast in the CST regime, resulting in all TCLs evolving around the control setpoint, the CPI regime reveals emergence of a \emph{bi-modal probability distribution and two, possibly strongly separated, time scales}. We observe that the two modes in the CPI regime are associated with all TCLs being in the same low and high-temperature states, respectively, with occasional (and therefore possibly rare) collective transition between the modes akin in the Kramer's phenomenon of statistical physics. To the best of our knowledge, this phenomenon was overlooked in the context of the multi-zone energy building engineering, even thought it has  direct implications on the operations of centralized cooling systems in buildings. It teaches us that a balance needs to be struck between occupational comfort --- related to variations in the individual temperatures --- and power output predictability --- the main focus of the DR schemes.
    
\end{abstract}

\maketitle

\section{Introduction}

The essence of Demand Response (DR) lies in providing auxiliary services helping power system operators manage uncertainty. The latter can emerge both from variable generation, e.g. as wind and solar, or from electricity market volatility~\cite{Palensky2011DemandLoads}. The key idea behind DR consists in leveraging flexible and inexpensive resources on the demand side of the power balance to ensure stability. The essential source of flexibility lies in the fact that many consumers of electricity can tolerate consumption delays, provided that some constraints remain satisfied~\cite{Oconnell2014BenefitsReview}. In addition to large and stable loads, aggregations of many small loads, e.g. residential appliances, can also be involved in DR services~\cite{Dhulst2015DemandBelgium}. The heating and cooling system in residential buildings is one such load that possesses inherent flexibility due to thermal inertia, and thereby presents significant opportunities in a DR market. Specifically, the potential for impacting total load via setpoint changes has been investigated numerically~\cite{Yin2016QuantifyingChanges, Finck2018QuantifyingSystems} and empirically~\cite{Xu2010DemandAudits, Kara2014QuantifyingApproach}, as well as its 
ramifications for occupant comfort~\cite{Aghniaey2018TheReview}. 

\subsection{Related Work}

Theoretical studies on the matter have essentially anchored around the so-called Thermostatically Controlled Loads (TCLs), denoting physical entities whose temperature oscillates within a range or around a target value; examples include rooms in building or refrigerators. Understanding the behavior of aggregations of individual entities and the underlying potential for DR has underpinned most of the interest in this discipline. 

Initial studies focused on adapting existing and developing new approaches to the statistics of TCLs \cite{Chong1979Statisticalmodels}, under different characteristics such as lifestyle and weather~\cite{Ihara1981PhysicallyPickup}, and proposed a methodology for the aggregation of individual loads \cite{Malhame1984StatisticalPickup}. Large aggregates, also called ensembles, were studied with the tools of statistical physics, such as Fokker-Planck equations~\cite{Malhame1985ElectricSystem, Malhame1988OnLoads}, and of control and reinforcement learning, such as Markov Decision Processes \cite{chertkov_ensemble_2017,18ChertkovDekaDvorkin}. 
The main operational philosophies in the literature include individual thermostat setpoints control~\cite{Callaway2009a, Callaway2011Achievingloads, Bashash2011ModelingLoads} or randomization and automatic feedback control at the individual TCL level, based on collective output~\cite{metivier_power_2019,Metivier2020Mean-fieldLoads}.

\begin{figure}
    \includegraphics[height = 5cm]{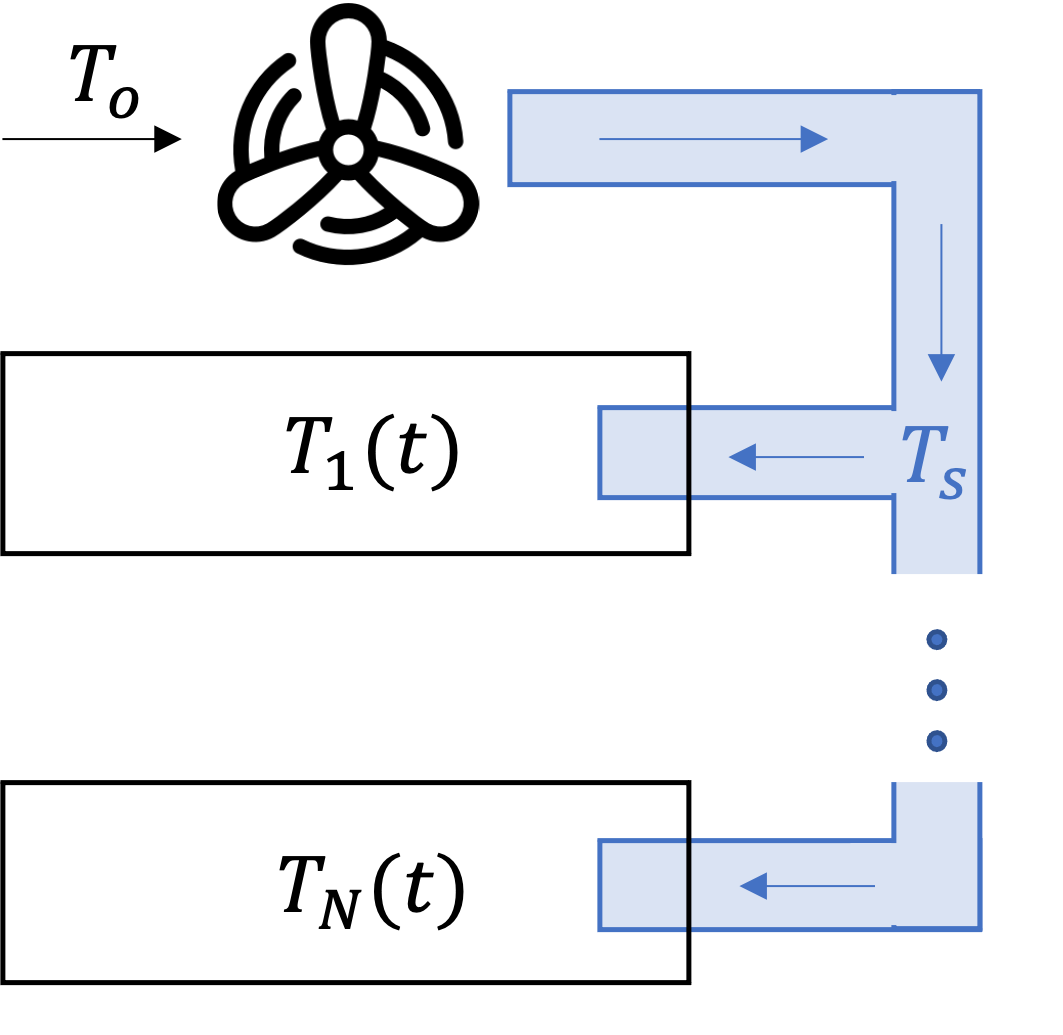}
    \caption{Simplified illustration of the setup considered. The AHU is represented by the fan, while individual units have their own temperatures. The exhaust system from each unit is not represented for simplicity, and air recirculation is neglected.}
    \label{fig:illustration}
\end{figure}

 While standard TCL models do represent some installations like \textit{independent} AC units, or refrigerators, they do not directly capture the intricate dynamics emerging from the coupling of those units via a district heating network~\cite{Claessens2018Model-freeNetwork}, or the grid services they provide \cite{Webborn2017AControl, Bashash2013ModelingManagement, Callaway2009a}.
In the context of heating and cooling within a multi-zone building, individual zones are thermally regulated by a small number of Air Handling Units (AHU), see Fig.~\ref{fig:illustration} for a simplified illustration. Each AHU is connected to some number of zones. Its role is to cool and dehumidify a mix of outside and recirculated air to a given temperature and relative humidity, and then to circulate this air --- at a given temperature, the \textit{supply} temperature --- throughout the building. The airflow is distributed to each zone via  \textit{Variable Air Volume} (VAV) boxes.
See \cite{Ma2012a} for an accurate and detailed description of the operations of such systems.

Standard TCL models fail to account for the specificities of such systems in two ways. First, they usually assume units to be independent. However, they are actually \textit{coupled} through AHU  -- that is adjusting to the supply temperature. This interaction and its implication for control has not been studied extensively. Second, the dynamics of cooling in these models is very different from the reality of residential and commercial buildings. Indeed, in standard TCL models, cycling is inherent in the dynamics. A unit is cooled at \textit{constant power} until it crosses the lower limit of a \textit{deadband}, and it is then left to reheat until it crosses its upper limit, at which point the cycle restarts. In AHU systems, such cycling is not enforced directly, and the cooling power changes following changes in the supply temperature~\cite{MacDonald2020ABuildings}. We will see below that the abscence of a prescribed \textit{deadband} has far-reaching consequences on how fast the system can recover from a perturbation, e.g. caused by the units following a DR request.

\subsection{Contribution}
In this manuscript, we suggest new models to represent TCLs coupled via an AHU, and study the thermalization dynamics of the system subject to two-level control -- at the collective (AHU) level and  at the individual (TCL) level. Specifically, we consider two principally different models representing distinct scenarios:
\paragraph{Constant Supply Temperature (CST)} The supply air temperature cooling each TCL is held constant over time, thereby eliminating direct coupling between individual units. While simplistic, this model enables the understanding of the fundamental dynamics of TCL units. 
\paragraph{Constant Power Input (CPI)} In order to reflect actual building operations, TCLs are coupled via the supply temperature. The latter is adjusted in order to ensure constant power input. 

In both cases, we focus on analyzing the details of the stabilization of the ensemble to a statistical steady state, established as a result of a balance between exogenous fluctuations representing TCL-specific perturbations and thermal relaxation. We combine analytical and numerical analyses to derive insights into the dynamics of this system. In the CST model, we find that thermalization is relatively fast, and results in TCLs stabilizing around the prescribed temperature. However, in the CPI scenario, we observe the appearance of more complex dynamics. When the level of stochastic fluctuations is small, all units stabilize in either of the two states associated with low-airflow and high-airflow.  Transitions between both modes, and thus relaxation to a statistically steady state, happens at a time scale which is much longer that the natural time scale of the problem, associated with meta-stable equilibration around one of the modes. Increasing the level of fluctuations results in a phase transition to an "entropic" state distributed around the control setpoint. We argue that this complex behavior of the CPI model is akin to the phenomenon of the thermally-activated barrier crossing in statistical mechanics, often referred to as the reaction-rate, or Kramers' theory \cite{1940Kramers,1990Hanggi}.

These observations are directly relevant for the operations of those common coupled systems. Under low uncertainty and perturbations, one can argue that operating the system at fixed supply temperature (CST) will result in a predictable outcome matching comfort requirements, with minor fluctuations in the total power required. However, in the case of higher fluctuations, e.g. associated with other uncertainties or driving forces such as sun irradiation, the dynamics of such a system is non-trivial and implies a need to establish a balance between comfort and requirements for utilization in DR.

\subsection{Outline}
The contents are presented as follows. In Section \ref{sec:formulation}, the problem is formulated along with the two scenarios of interest. A statistical analysis of the CST scenario is conducted in Section~\ref{sec:CST}. Analyses under Constant Power Input are conducted in Section~\ref{sec:CPI}. We first consider the noiseless limit for 1 TCL, and extend its insights to the presence of noise. The general case with an arbitrary number of TCLs is studied in Section~\ref{sec:many}. We conclude and discuss the path forward in Section~\ref{sec:conclusions}.

\section{Problem Formulation}\label{sec:formulation}

The system we consider consists of two levels. At the lower level, TCLs evolve according to their individual dynamics. At the higher level, the Air Handling Unit manages the energetic input to all TCLs, thereby coupling them to one another (cf Fig.~\ref{fig:illustration}). In this Section, we lay out the models governing both levels.

\subsection{TCL dynamics}

We model the dynamics of an individual TCL by the following Stochastic Differential Equation (SDE): 
\begin{gather}
c\ dT_i =-f(T_i(t);s(T_i(t));T_s)\ dt+\sqrt{2D}\ dW_i(t),\label{eq:Ti}\\
\begin{split}
f(T_i(t);s(T_i(t));T_s) &= \frac{T_i(t)-T_o}{r} \\
&+\bar{\mu}s(T_i(t)) c_p\left(T_i(t)-T_s\right),\label{eq:Ti2}\\
\end{split}\\
s(T_i(t))=\left\{\begin{array}{ll} s_-, & T_i(t)\leq \bar{T}\\ s_+, & T_i(t)> \bar{T}\end{array}\right..\label{eq:Ti3}
\end{gather}
where $t$ is time and $T_i(t)$ is the temperature of the $i$-th TCL unit $i\in 1, ..., N$. We assume that all units are identical, and characterized by the following physical parameters: $r\ [K/kW]$ is the coefficient of thermal resistance to the outside air; $c_p\ [kJ/kgK]$ is the specific heat capacity of air; $c\ [kJ/K]$ is the capacitance coefficient representing the total air-mass associated with an individual TCL; $W_i(t)$ is the Wiener process; and $D\ [kJ^2]$ represents the amplitude of exogenous, stochastic, zero mean thermal (white) noise affecting the TCLs (associated, e.g., with some uncertainty around in-zone traffic and operations). $\bar{\mu}$ denotes the maximum airflow that can get into the zone, while $s(\cdot)\in [0, 1]$ indicates how much air is actually flowing into the room at time $t$.
In Eq.~\eqref{eq:Ti2}, describing the temperature dynamics of a TCL unit, $f(T_i(t);s(T_i(t));T_s)$ denotes the thermal force associated with two principally different terms. The first contribution to the thermal force describes deterministic relaxation of the zone temperature, $T_i(t)$, with the rate $r$ to the ambient temperature $T_o$, which is taken as constant in this document. The second contribution to the thermal force represents the injection of cool air delivered by VAV units at the mass flow rate $\bar{\mu}s(T_i(t))$ and at a temperature equal to the supply temperature $T_s$ into the $i$-th zone. In the present system, we assume the VAVs are able to modify only the airflow entering each zone. While in general they are also able to reheat the incoming air, we assume this is not applicable, as is the case in warm regions where cooling is mostly needed.
The dynamics of the air flow at each VAV is a simplification of the dual-maximum control logic~\cite{Taylor2012DualLogic}: we neglect the heating regime and we assume instantaneous switching as the setpoint is crossed instead of progressive changes of the airflow (Eq.~\eqref{eq:Ti3}). The model proposed is a degenerate case of the more general one, with the benefit of simplicity and tractability.

\subsection{AHU dynamics}

The model in Eqs~\eqref{eq:Ti}-\eqref{eq:Ti3} is incomplete until one provides a closure relationship for $T_s$, controlling the second contribution to the thermal force in Eq.~(\ref{eq:Ti2}). As detailed in the remainder of this Section, we consider two scenarios which will be different in terms of the relation between $T_s$ and the vector of TCL temperatures, ${\bm T}=(T_i|i=1,\cdots,N)$. In this paper, we consider simplified models of the AHU energy consumption. Indeed, we assume that the units lie in areas where the energy attributed to cooling dominates that associated to de-humidification. This means that the energy consumption of the AHU can be modeled by the enthalpy change of the air as it goes through the AHU~\cite{Sterling2014Model-basedMethodologies, Razban2019ModellingBuilding, Zajic2011ModellingApproach,Afram2014ReviewSystems}. For the sake of simplicity, we also neglect air recirculation within the building, and assume that 
only the outside air is used as input into the AHU.

\subsection*{(a) Constant Supply Temperature (CST)}

The first scenario considered, indexed by $a$, consists in fixing the temperature $T_s$ of the air delivered to the TCLs to a prescribed constant value, $T_{s;a}$. The total power, i.e. energy injected into the ensemble per unit time, then writes
\begin{gather} \label{eq:Pa}
    P_a(t)=\mu(t)\left(T_o-T_{s;a}\right),
    \quad \mu(t) = \bar{\mu}\sum_i s(T_i(t)), 
\end{gather}
where $\mu(t)$ is the aggregate airflow within the AHU. We assume that all the airflows in the system are balanced at any moment of time. Notice that under this CST scenario, where $T_s=\text{const}$, $s(T_i(t))$ changes with time due to the evolution of $T_i$, thereby driving changes in $\mu(t)$. This results in a change of the total power consumed by the ensemble, $P(t)$, thereby inheriting some temporal dynamics.

\subsection*{(b) Constant  Power Input  (CPI)}

In this second scenario, indexed by $b$, we assume that the system operator keeps the power input, $P$, constant at all times, i.e. $P=P_b=\text{const}$. This is achieved by adjusting the supply temperature, $T_s(t)$, according to the following modification of Eq.~(\ref{eq:Pa}):
\begin{gather}\label{eq:Tb}
    T_{s;b}({\bm T}(t))=T_o-\frac{{P}_b}{\bar{\mu}\sum_i s(T_i(t))}.
\end{gather}
We see that the supply temperature is continually adjusted based on the local temperature dynamics.

\section{Statistical Analysis of the CST Scenario}
\label{sec:CST}

The supply temperature $T_s$ is assumed constant, $T_s=T_{s;a}$. The system of SDEs \eqref{eq:Ti}-\eqref{eq:Ti3} is closed, thus translating into the so-called Kolmogorov-Fokker-Planck (KPF) \cite{2004Gardiner,2007vankampen} Partial Differential Equation (PDE) for the joint probability distribution of the vector of temperatures ${\bm T}$ within the TCL ensemble: 
\begin{gather}\label{eq:KFP-a}
  \hspace{-0.3cm}\partial_t {\cal P}_a({\bm T}|t)= \frac{1}{c} \sum_i\ \partial_{T_i} \left( f(T_i;s(T_i);T_{s;a}) 
  +D\partial_{T_i} 
  \right){\cal P}_a({\bm T}|t).
\end{gather}
We observe that the KFP Eq.~(\ref{eq:KFP-a}) can be represented in the potential form, 
\begin{eqnarray}\label{eq:KFP-a-J}
    \hspace{-0.4cm}\partial_t {\cal P}_a({\bm T}|t) &=&-\sum_i\partial_{T_i} J_i,\\ \label{eq:J-a}
    \hspace{-0.4cm} J_i & = & - \frac{1}{c}\left(\partial_{T_i} U(\textbf{T}|T_{s;a}) +D \partial_{T_i}\right) {\cal P}_a({\bm T}|t),\\ \label{eq:U}
    \hspace{-0.4cm} U({\bm T}|T_{s;a}) &=& \sum_{i=1}^N U_1(T_i|T_{s;a}),\\ \label{eq:U1}
    \hspace{-0.4cm} U_1(T_i|T_{s;a}) &=& \frac{(T_i-T_o)^2}{2r}\\\nonumber &+&\frac{c_p \bar{\mu} s(T_i)}{2}\left((T_i-T_{s;a})^2 - (\bar{T}-T_{s;a})^2\right),
\end{eqnarray}
where  $J_i$ is the probability current along $T_i$, and $U({\bm T}|T_{s;a})$ and $U_1(T_i|T_{s;a})$ are the aggregated and individual TCL thermal potentials, respectively. 
Notice that the KFP Eq.~(\ref{eq:KFP-a}) describes a Cauchy, i.e. an initial-value, problem. It should therefore be equipped with an initial condition, e.g. that all TCLs are at the same temperature $T_0$ at $t=0$. Then, ${\cal P}_a(T|0)=\prod_i\delta(T_i-T_0)$, where $\delta(\cdot)$ is the Dirac $\delta$-function. More generally, if ${\cal P}_a({\bm T}|0)$ is factorized into a product of marginal distributions of individual $T_i$, the solution to the KFP equations at all $t$ is also factorized into the product of the corresponding marginal probability distributions, i.e. ${\cal P}_a({\bm T}|t)=\prod_i {\cal P}_{a;1}(T_i|t)$.

\subsection{Stationary Distribution} 

Due to the factorized structure of the differential operator on the right hand side of Eq.~(\ref{eq:KFP-a}), the steady-state solution to Eq.~(\ref{eq:KFP-a}), ${\cal P}_{a;st}({\bm T})$, i.e. one acquired at $t\to\infty$, is also factorized into the product of the respective marginals, ${\cal P}_{a;st}({\bm T})=\prod_i {\cal P}_{a;1;st}(T_i|T_{s;a})$. Moreover, ${\cal P}_{a;1;st}(T_i|T_{s;a})$ satisfies a second order ODE which can be solved for any values of $T_{s;a}$, considered as a parameter: 

\begin{align}\label{eq:a-Pst}
   {\cal P}_{a;1;st}(T_i|T_{s;a})&= \frac{1}{Z}\exp\left(-\frac{cU_1(T_i|T_{s;a})}{D}\right),\\
    \label{eq:a-Zst} 
      Z&= \int\limits_{-\infty}^{\infty}\!\!dT \exp\left(-\frac{c U_1(T|T_{s;a})}{D}\right).
\end{align}

 The single-TCL thermal potential, $U_1(T_i|T_{s;a})$, defined in Eq.~(\ref{eq:U}), is continuous in $T_i$. It is however not smooth as the derivative jumps at $T_i = \bar{T}$, and attains a single minimum at
\begin{gather}\label{eq:T-a-min}
    \hspace{-0.4cm}T_{a;min}(T_{s;a})=\left\{\begin{array}{ll} \frac{T_o+T_{s;a}\mu^{(-)} c r}{1+\mu^{(-)} c r},& \hspace{-0.1cm}T_{s;a}\leq \beta^{(-)};\\
    \bar{T},& \hspace{-0.1cm}\beta^{(-)}\leq T_{s;a}\leq \beta^{(+)};\\
    \frac{T_o+T_{s;a}\mu^{(+)} c r}{1+\mu^{(+)} c r},&  \hspace{-0.1cm}\beta^{(+)}\leq T_{s;a}.\end{array}\right.
\end{gather}
where $\beta^{(\pm)} = \bar{T}- (T_o-\bar{T})/(\mu^{(\pm)}cr)$. 
The most probable value of the TCL temperature in the steady state does not depend on the amplitude of the thermal noise, $D$.

\begin{figure}
    \centering
    \includegraphics[width=.4\textwidth]{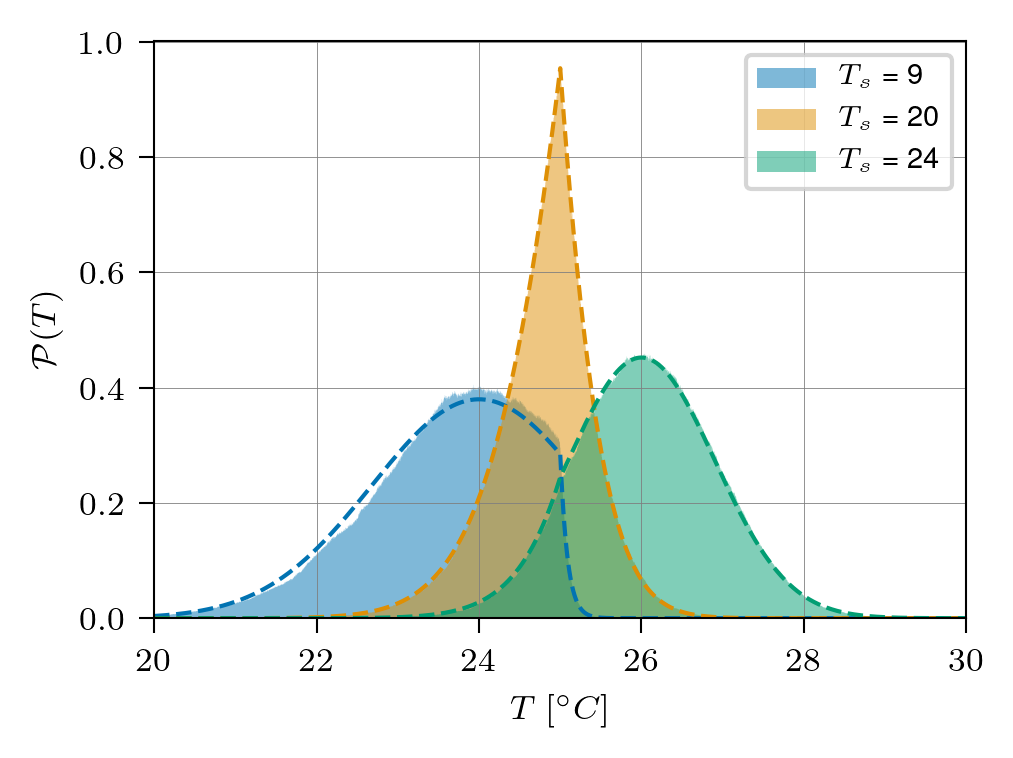}
    \caption{Temperature distributions for an individual TCL under the CST scenario after initial transients, for different values of the supply temperature $T_s$. $c = 20.5\ kJ/K$. \textbf{Filled:} Experimental results. \textbf{Dotted lines:} Theoretical distributions obtained according to Eq.~\eqref{eq:a-Pst}.}
    \label{fig:Ui-sep}
\end{figure}

We report in Fig.~\ref{fig:Ui-sep} the dependence of the steady-state individual TCL probability distribution ${\cal P}_{a;1;st}(T_i|T_{s;a})$ on $T_i$ at different values of $T_{s;a}$
\footnote{Unless noted otherwise, parameters for numerical results are: $T_o=30^\circ C$, $s_- = 0.2$, $s_+=1$, $\bar{\mu} = 1\ kg/s$, $r=2\ K/kW$, $c = 15\ kJ/K$, $\bar{T} = 25^\circ C$. 
}.
Note that there is a range of supply temperature values where the most probable value is achieved at $T_i=\bar{T}$, e.g. for $T_s = 20^\circ C$ in Fig.~\ref{fig:Ui-sep}. This maximum appears because fluctuations drive the temperature across $\bar{T}$, thereby implying repetitive switching of the airflow $s(T_i)$, and on average stabilizing the temperature around the setpoint.

\subsection{Relaxation to the Steady Distribution}

Even in the case when the initial probability distribution cannot be factorized into a product of independent terms, independence will appear dynamically at sufficiently large $t\gg \tau_a$,  where $\tau_a$ is the so-called mixing time\footnote{$\tau_a$ is defined as a  time when ${\cal P}_{a;1}(T|t;T_{s;a})$ gets sufficiently close, wrt norm $L_2$ or Kullback-Leibler divergence, to its steady asymptotics, i.e. $\|{\cal P}_{a;1;st}(T|T_{s;a})-{\cal P}_{a;1}(T|t;T_{s;a})\|\propto \exp(-t/\tau_a) \ll 1$}. 
One can extract $\tau_a$ and, specifically, its dependence on the supply temperature $T_{s;a}$ from the spectral analysis of the the $i$-th component of the differential operator entering the KFP Eq.~(\ref{eq:KFP-a}). 
The resulting dependence does not hold any specific and unexpected features, and is therefore not reported here.

\section{Statistical Analysis of the CPI Scenario}
\label{sec:CPI}

The most significant difference between the CST and the CPI scenarios lies in the stochastic coupling between individual TCLs. According to Eq.~\eqref{eq:Tb}, the supply temperature $T_s$ directly depends on all TCL temperatures. In this Section, we discuss the significance of this modification.

Informally, we expect that this case is special because the stochastic dynamics of individual TCLs are no longer decoupled. 
General theory of KFP equations suggests \cite{2004Gardiner,2007vankampen} that at $N>1$ in the CPI scenario (a) the thermal force can no longer be represented as a gradient of a potential; (b) detailed balance is broken; (c) the steady distribution is no longer a Gibbs distribution and cannot be factorized into a product of components, each representing an individual TCL. 

In this Section, we provide a quantitative analysis of the aforementioned expected qualitative behavior of the system. In Section~\ref{sec:D=0}, we analyze the system in the noiseless limit, i.e. $D=0$, and show that the system is strongly sensitive to initial conditions via the identification of two different fixed points governing the long-time behavior of the system. In Section~\ref{sec:one}, we reintroduce noise in the context of a single TCL as it enables some analytical treatment. We leverage the latter in Section~\ref{sec:many} where the general case of $N>1$ TCLs with noise is considered and where we discover that several features from the $N=1$ case surprisingly hold.

\subsection{Noiseless Limit}\label{sec:D=0}

\begin{figure}
    \centering
    \includegraphics[scale=1]{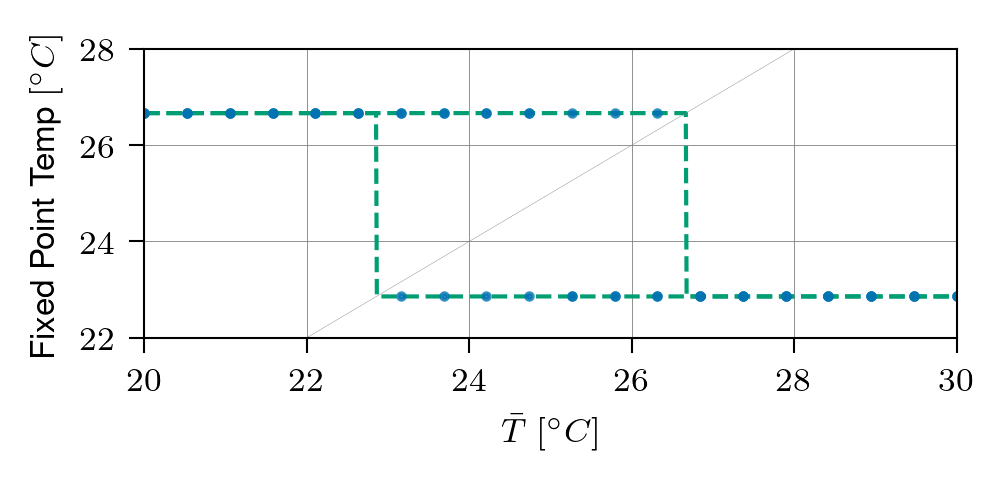}
    \caption{Location of the two fixed points for $N=2$ in the CPI scenario. The identity is represented by the gray diagonal line. 
    \textbf{Green line, dashed:} Expected position of the two fixed points according to Eq.~\eqref{eq:Tb-1}.
    \textbf{Blue dots:} Final temperatures of both TCLs attained after $100 \unit{s}$ under multiple initializations.}
    \label{fig:b_bifurcation}
\end{figure}

In order to extract generic statements about the behavior of this coupled system, we analyze the fixed points of Eq.~\eqref{eq:Ti} with $T_s$ substituted by $T_{s;b}({\bm T}(t))$ from Eq.~\eqref{eq:Tb} in the noiseless case, i.e. in the deterministic regime.
As TCLs are indistinguishable, a fixed point solution is determined by the number of TCLs in the high-airflow regime, $N_+(t)=\sum_i \theta(T_i(t)>\bar{T})$, where $\theta(\cdot)$ is the indicator function. We observe that there are potentially $N+1$ fixed point solutions to Eq.~\eqref{eq:Ti} complemented with the power balance in the CPI scenario, i.e. $N_+(t) \in \{0,\dots,N\}$. According to Eq.~\eqref{eq:Tb}, $T_{s;b}$ is directly determined by $N_+(t)$: $T_{s;b}(N_+(t)) = T_o - P_b/\bar{\mu} (N_+s_+ + (N-N_+)s_-)$. 
For each of the $N+1$ possible states, the $N$ TCLs are split into two groups: a group of $N_+$ units in the high-flow regime and another $N-N_+$ in the low-flow regime. In addition, in the deterministic case, all TCLs in a given group are at the same temperature at steady-state, given by
\begin{gather}\label{eq:Tb-1}
T_{\pm}(N_+(t))=\frac{T_o+T_{s;b}(N_+(t))\mu_{\pm} c r}{1+\mu_{\pm} c r}.
\end{gather}
An immediate consequence of this equation is that $T_+(N_+(t))<T_-(N_+(t))$, providing that $\mu_+>\mu_-$. However, this is in contradiction with the requirement that $T_+(N_+(t))>\bar{T}>T_-(N_+(t))$.
In other words, there can be no stable fixed points, i.e. states realized dynamically at $t\to \infty$ in the noiseless regime, with a nonzero number of TCLs in both states (high-flow and low-flow). Therefore two options are left for stable fixed points: $N_+(t)=0$ or $N_+(t)=N$. Which of the two stable fixed points is attained at long times will depend on the value of $\bar{T}$. Indeed, the $N_+(t)=0$ state is valid (self-consistent) if
$T_-(N_+(t)=0)<\bar{T}$ and the $N_+(t)=N$ state is valid if $T_+(N_+(t))>\bar{T}$. In addition, $T_-(N_+(t)=0) < T_+(N_+(t)=N)$. Therefore, the steady state that will be attained will consist of (i) $N_+(t)=N$ if $\bar{T}<T_-(N_+(t)=0)<T_+(N_+(t)=N)$, (ii) $N_+(t)=0$, if $T_-(N_+(t)=0)< T_+(N_+(t)=N)<\bar{T}$, or (iii) either of the two if $T_-(N_+(t)=0)<\bar{T}<T_+(N_+(t)=N)$, in which case the final state is directly dictated by the initial conditions $T(0)$. Fig.~\ref{fig:b_bifurcation} illustrates the dependence of the realized fixed points on the values of $\bar{T}$ for different initial conditions in the case of $N=2$ TCLs. $T_\pm$ do indeed act as stable fixed points in the noiseless limit. Depending on $\bar{T}$, either one or both of them are accessible to the system. 

\subsection{Case of $N=1$, $D>0$.}\label{sec:one}
\begin{figure}
     \centering
     \begin{subfigure}[b]{.4\textwidth}
         \centering
         \includegraphics[width=\textwidth]{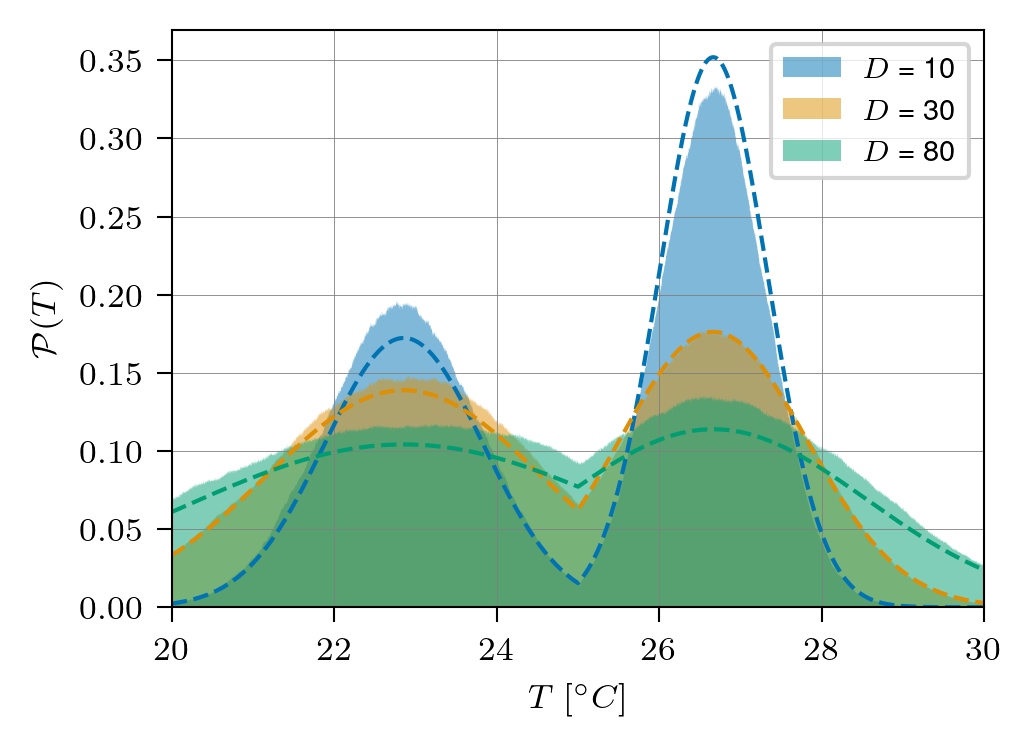}
         \caption{Temperature distribution of 1 TCL under the CPI scenario after initial transients, for different values of the noise amplitude $D_i$. \textbf{Filled:} Experimental results. \textbf{Dashed line:} Theoretical distributions obtained according to Eq.~\eqref{eq:a-Pst}. 
        }
         \label{fig:b_pdf_1}
     \end{subfigure}
     \\
      \begin{subfigure}[b]{.4\textwidth}
         \centering
         \includegraphics[width=\textwidth]{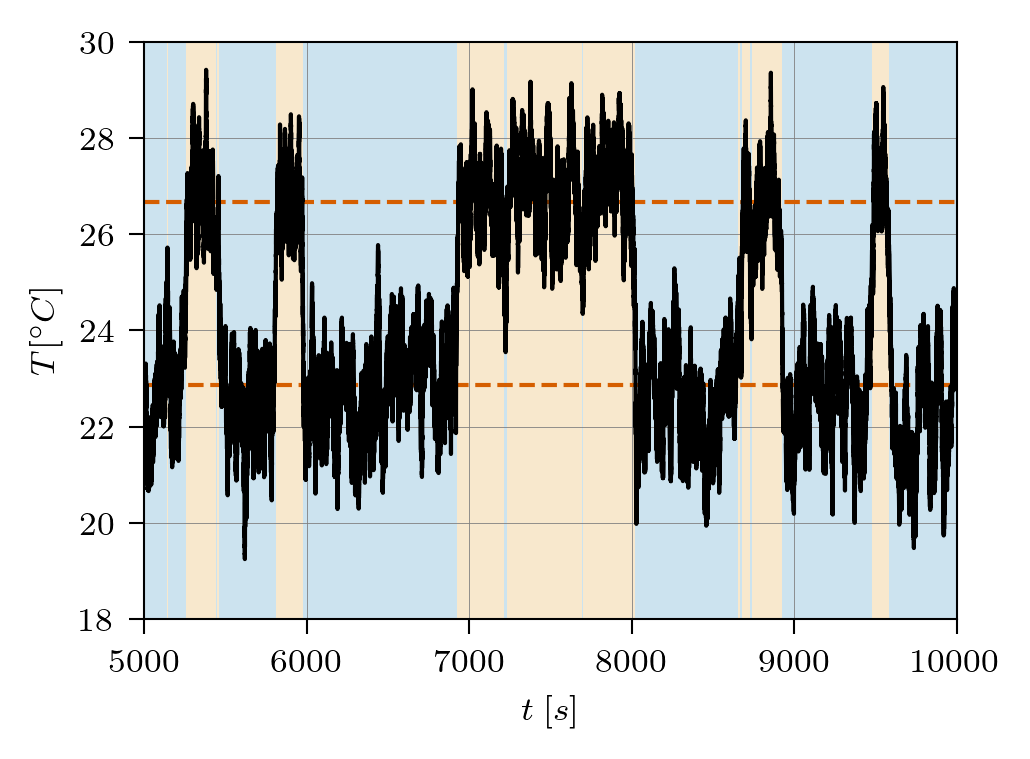}
         \caption{Time series of 1 TCL under the CPI scenario, with $D_i = 15\ kJ^2$. \textbf{Red lines, dashed:} Locations of the theoretical fixed points $T_\pm$ (cf. Eq.~\eqref{eq:Tb-1}). \textbf{Filled background:} A change in background color indicates a change in the value of $T_{min}(T_s)$.}
         \label{fig:b_time_series_1}
     \end{subfigure}
     \caption{Dynamics and steady state distribution of 1 TCL under the CPI scenario.}
     \label{fig:1TCL}
\end{figure}

As analysis in the general case with $N>1,\ D>0$ is difficult, we here focus on the stochastic case for 1 TCL. Eqs~\eqref{eq:U1},~\eqref{eq:a-Pst} still hold in this context, with the difference that the supply temperature is dictated by Eq.~\eqref{eq:Tb}. 
A similar analysis as that conducted in Section~\ref{sec:CST} is hence possible, and results are reported in Fig.~\ref{fig:1TCL}.

In contrast to the single-peak distributions observed in the CST scenario (cf. Fig.~\ref{fig:Ui-sep}), one observes the emergence of two maxima corresponding with the two fixed points identified in Section~\ref{sec:D=0}. The position of these stable temperatures is directly determined by the physical parameters of the system, while their relative importance can be tuned by the setpoint $\bar{T}$.
We attribute the mismatch between analytical and empirical distributions in Fig.~\ref{fig:b_pdf_1} to a finite computational budget and step size. The stochastic nature of the system actually enables switching between both modes. As illustrated in Fig.~\ref{fig:b_time_series_1}, a single TCL driven by noise successively switches between the two equilibrium points --- thereby implying an adaptation of the supply air temperature $T_s$, not shown here. 

As the potential $U_1(T)$ has two distinct minima in some regimes, Kramers' theory \cite{1940Kramers,1990Hanggi} suggests the emergence of two distinct temporal scales. One short time scale $\tau_{s}$ should be associated to equilibrium fluctuations around either of the two minima, while the transition between either minimum as seen in Fig.~\ref{fig:b_time_series_1} is characterized by a longer one $\tau_{l}$. These are related via $\tau_l \sim \tau_s \exp(\Delta/D)$, where $\Delta$ is the potential barrier at $T_1 = \bar{T}$. Strong separation of these timescales, i.e. $\tau_l \gg\tau_s$, emerges when $D\ll\Delta$.

\subsection{Numerical Experiments for $N>1, D>0.$}\label{sec:many}

Even if, strictly speaking, insights from Kramers' theory only apply to the $N=1$ case discussed above, most qualitative features seem to translate to the general $N>1$ case. 

\begin{figure}
     \centering
     \begin{subfigure}[b]{.4\textwidth}
         \centering
         \includegraphics[width=\textwidth]{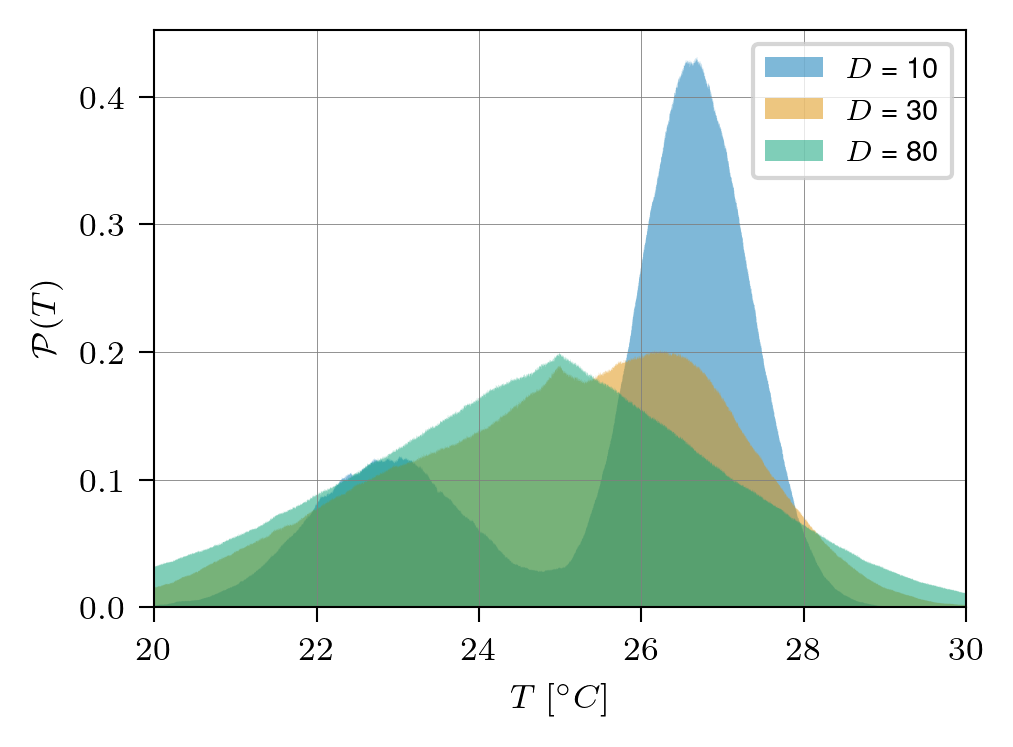}
         \caption{Empirical temperature distributions of TCLs under the CPI scenario with $N=5$, after initial transients and for different values of the noise amplitude $D_i$.}
         \label{fig:b_pdf_5}
     \end{subfigure}
     \\
     \begin{subfigure}[b]{.4\textwidth}
         \centering
         \includegraphics[width=\textwidth]{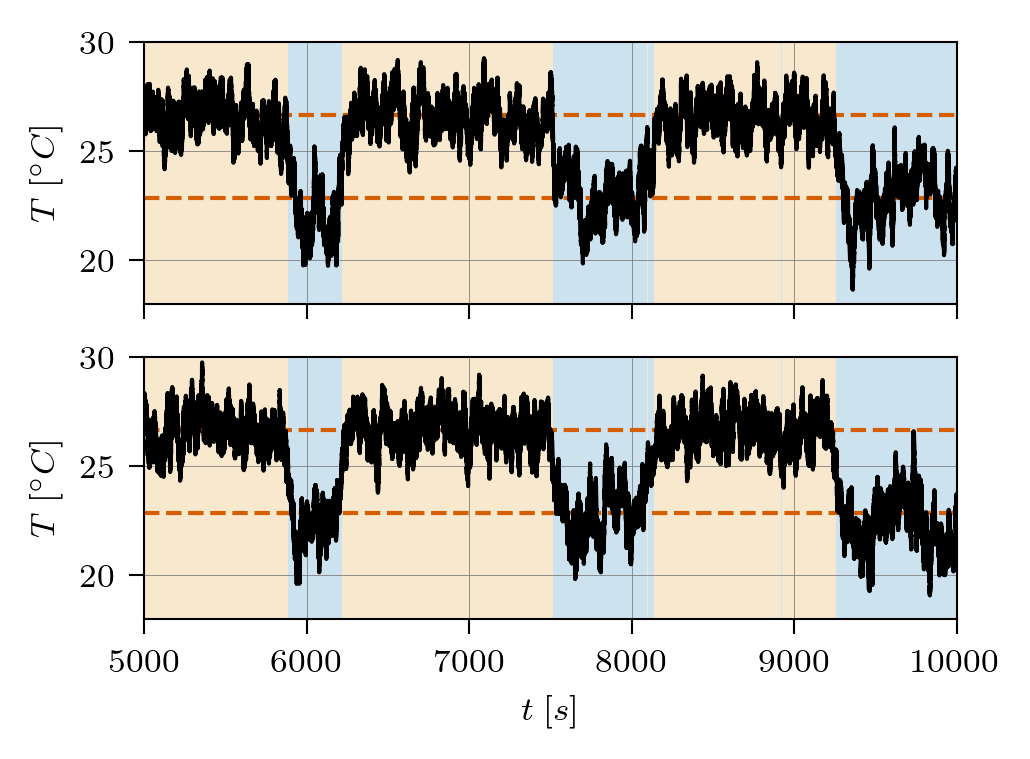}
         \caption{Time series of 2 TCLs under the CPI scenario, with $N=2$. \textbf{Red lines, dashed:} Locations of the theoretical fixed points $T_\pm$. \textbf{Filled background:} A change in background color indicates a switch in the value of $T_{min}(T_s)$.}
         \label{fig:b_time_series_N}
     \end{subfigure}
     \caption{Temperature distribution and dynamics in the general case of $N>1, D>0$ CPI case.}
     \label{fig:NTCLs}
\end{figure}

\begin{figure}[ht]
     \centering
     \begin{subfigure}[b]{.4\textwidth}
         \centering
         \includegraphics[width=\textwidth]{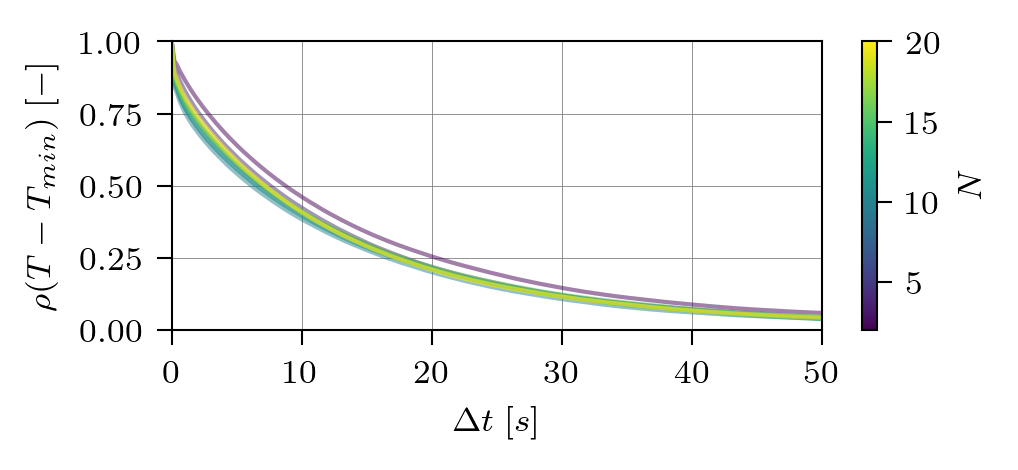}
         \caption{Autocorrelation of $T(t)-T_{min}(t)$. The exponential decay is indicative of a short associated timescale.}
         \label{fig:autocorr-t}
     \end{subfigure}
     \\
     \begin{subfigure}[b]{.4\textwidth}
         \centering
         \includegraphics[width=\textwidth]{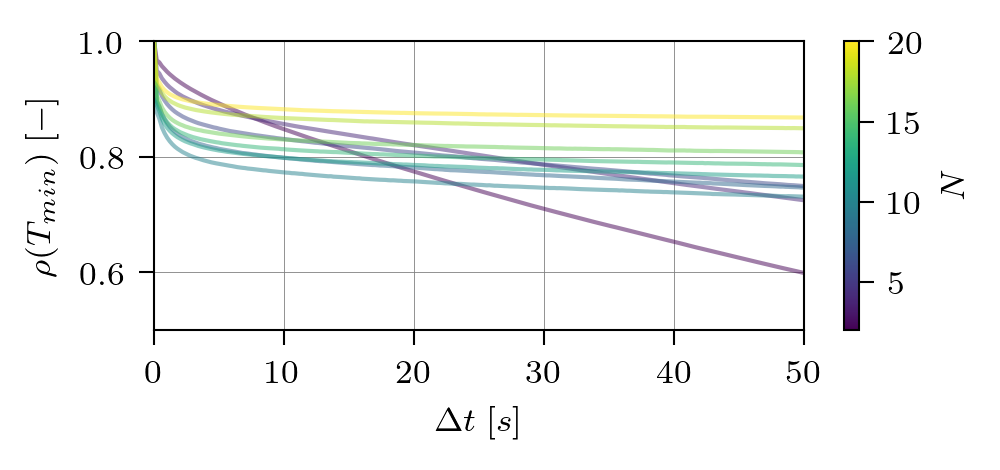}
         \caption{Autocorrelation of $T_{min}$, defined in ~\eqref{eq:T-a-min}. The decay is much slower than for Fig.~\ref{fig:autocorr-t}, and the characteristic timescale increases with the number of TCLs in the system.
         }
         \label{fig:autocorr-tmin}
     \end{subfigure}
     \caption{Statistical analysis and observation of two different timescales in the system.}
     \label{fig:autocorrelation}
\end{figure}

\begin{figure}[ht]
    \centering
    \includegraphics[width = .4\textwidth]{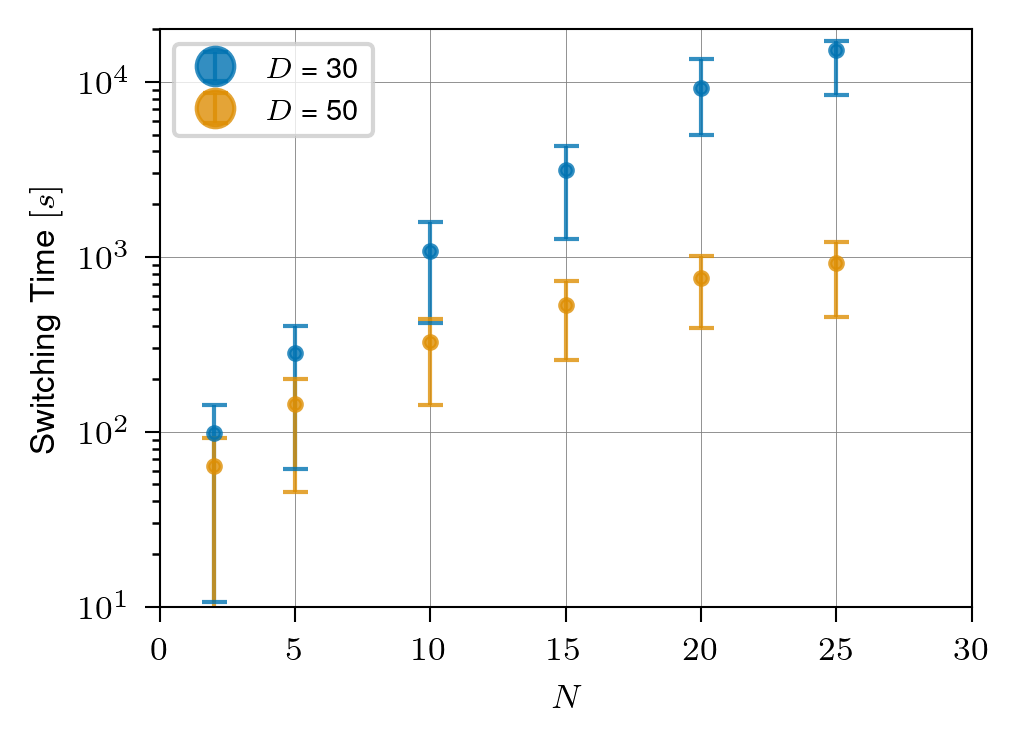}
    \caption{Scaling of the switching time with the number $N$ of TCLs in the CPI scenario. Dots indicate the average switching time. Errorbars indicate the 75th and 25th percentile of the distributions. }
    \label{fig:switching_time_vs_N}
\end{figure}

\paragraph{Bi-modal dynamics and emergence of two timescales} Up to moderate values of the noise amplitude $D$ a similar bi-modal distribution of accessed temperatures is observed (cf. Fig.~\ref{fig:b_pdf_5}, with $D = 10\ kJ^2$). Oscillations of temperatures around two central values are clearly visible in Fig.~\ref{fig:b_time_series_N}. Note that both TCLs illustrated exhibit coordinated switching from one fixed point to the other, resulting from the AHU coupling. Both observations are qualitatively analogous to the $N=1$ case, cf. Fig.~\ref{fig:1TCL}.  Specifically, one can also expect switching and the natural appearance of two characteristic timescales in the system. As shown in Fig.~\ref{fig:autocorr-t}, the autocorrelation $\rho(T(t)-T_{min}(t))$ of a given TCL temperature decays exponentially fast. 
However, the decay of $\rho(T_{min}(t))$, associated with the $N$ TCLs switching from one stable point to the next, is significantly slower in Fig.~\ref{fig:autocorr-tmin}. This suggests that memory is indeed present in the system and that switching is a rarer event with increasing values of $N$. Fig.~\ref{fig:switching_time_vs_N} provides another illustration of this strong $N$-dependence. The switching time over trials increases  with $N$, confirming the intuition provided by Fig.~\ref{fig:autocorr-tmin}. While the dependence of the switching time with $N$ seems sub-exponential, those results were obtained with a finite computational budget and might not reflect the true scaling. Further investigation is needed, via alternative theoretical and computational methods, to resolve it.

\paragraph{Emergence of an entropic state} Large noise amplitudes will prevent the system from settling in either of the stable points. Indeed, the two modes of the temperature distribution disappear as $D$ increases (see Fig.~\ref{fig:b_pdf_5}). Instead, they are slowly replaced by an \textit{entropic} peak at the location of the setpoint $\bar{T}$. From the perspective of a single TCL, this entropic state represents a balance between stochastic and deterministic forces resulting in the TCL switching back and forth between low- and high-flow regimes, and hence meandering around $\bar{T}$. In high noise regimes, the effect of the potential barrier disappears and the resulting distribution is akin in nature to the CST case in some regimes as shown in Fig.~\ref{fig:Ui-sep}. However, the noise intensity required to reach such regimes is such that the distribution broadens significantly with potential impact on occupants. 

\section{Conclusions and Path Forward}\label{sec:conclusions}

In this manuscript, we formulate a novel model describing the stochastic dynamics of coupled TCLs. It differs from standard TCL models in that it more accurately replicates cooling via a centralized Air Handling Unit. The main consequence of this modification is the resulting coupling between individual units, which is usually disregarded in other models. 

We analyzed the aggregate dynamics in two different scenarios. First, we considered the uncoupled dynamics of each unit, by fixing the supply air temperature (CST regime). While the CST regime is highly stable and predictable, fluctuations of the aggregated power consumption of the building, associated with airflow adjustments in each unit, might be undesirable. In the second CPI scenario, the power input is set constant, leading to significant coupling between individual TCLs and revealing interesting collective phenomena, with ramifications for the use of the ensemble in DR. The system's response to a perturbation may result in the co-existence of two quasi-steady modes associated with stable fixed points of the dynamics in the low-noise regime, which transition to an \textit{entropic} state when the level of the noise increases. Dynamic consequence of this complex behavior is the emergence of long, and generally parasitic, transients associated with the transitions between the two modes in the low-noise regime.

This new model and associated results set the stage for several future challenges:
\begin{enumerate}
\item Investigating the effects associated with inhomogeneity (disorder) in physical parameters, thereby describing populations of non-identical units, would provide a more accurate picture of real-life systems.

\item Developing methods leveraging both unit-level and system-level control to strike a balance between individual comfort and collective benefits would further our understanding of the inherent flexibility in such systems, and potentially result in innovative solutions for AHU control --- e.g. to remove the anomalously long relaxation times observed in the low-noise CPI regime. Ideas in recent papers~\cite{metivier_power_2019,Metivier2020Mean-fieldLoads} could be translated to the present case. Specifically, understanding the non-equilibrium nature of the TCL ensemble in the CPI regime and its practical significance for operations and control will be interesting. While similar to typical Kramers phenomena, the dynamics in the CPI regime are not at statistical equilibrium. Detailed balance is broken and fluxes emerge. The implications of such complex dynamics on controllability remain to be explored.

\item Confronting this new model with real zone-level time series and experiments~\cite{Keskar2018StaySystems, Li2015SimulationSystem} will provide crucial insight for its validity and highlight where improvements, modifications or relaxing assumptions are needed.

\item Analyzing the integration of such systems into larger-scale DR schemes, e.g. in the spirit of \cite{19Hasan,20Hasan}, and involving both multiple buildings as well as other appliances, will be critical in the deployment of such solutions.

\end{enumerate}

\section{Acknowledgments} 
LFV acknowledges Jacques de Chalendar for inspiring and helpful discussions, as well as for proof-reading the manuscript. Research Experience for Undegraduates work of LW at UArizona during the summer of 2021 was funded by NSF \# 1937229 "RTG: Applied Mathematics and Statistics for Data-Driven Discovery". 

\bibliography{bib/referencesLucas1,bib/TCLdisorder,bib/AdvancedTCL}

\end{document}